\documentclass[
]{arxiv}

\usepackage{listings}
\definecolor{figblue}{HTML}{3B76B1}
\definecolor{figorange}{HTML}{F09937}
\definecolor{dkgreen}{rgb}{0,0.6,0}
\definecolor{gray}{rgb}{0.5,0.5,0.5}
\definecolor{mauve}{rgb}{0.58,0,0.82}
\definecolor{PROVblue}{HTML}{CFCEFB}
\definecolor{PROVyellow}{HTML}{FFFFC6}
\definecolor{PROVorange}{HTML}{FCECCA}
\usepackage{hyperref}
\usepackage{amsmath,amsfonts}
\usepackage{todonotes}
\usepackage{paralist}

\definecolor{greenFig}{HTML}{4F8F00}
\definecolor{redFig}{HTML}{B51700}
\definecolor{blueFig}{HTML}{3274B4}

\newcommand{\pq}[1]{\textit{\textbf{#1}}}

\begin{document}

\title{Towards dimensions and granularity in a unified workflow and data provenance framework}

\tnotemark[1]
\tnotetext[1]{This is a preprint of a workshop paper, published at LWDA 2024.}

\author[1]{Tanja Auge}
\address[1]{University of Regensburg, Germany}

\author[2,3]{Sascha Genehr}
\address[2]{University of Rostock, Germany}
\address[3]{Rostock University Library, Germany}

\author[1]{Meike Klettke}

\author[4]{Frank Krüger}
\address[4]{Wismar University of Applied Sciences, Germany}

\author[3]{Max Schröder}

\begin{abstract}
Provenance information are essential for the traceability of scientific studies or experiments and thus crucial for ensuring the credibility and reproducibility of research findings. 
This paper discusses a comprehensive provenance framework combining the two types \begin{inparaenum}
    \item \textit{workflow provenance}, and 
    \item \textit{data provenance}
\end{inparaenum} as well as their dimensions and granularity, which enables the answering of \textbf{W7+1} provenance questions.
We demonstrate the applicability by employing a biomedical research use case, that can be easily transferred into other scientific fields.
An integration of these concepts into a unified framework enables credibility and reproducibility of the research findings.
\end{abstract}


\maketitle

\section{Introduction}
\label{sec:introduction}
The traceability of scientific studies and experiments is essential for the credibility and reproducibility of their findings.
Provenance information provide the essential data for this purpose, e.g.\,the sequence of activities that resulted in the creation of measurement data, the involved persons or the investigated samples.
Thereby, provenance can have a significant impact on the scientific value of the research data.
Many approaches and definitions of provenance in different contexts have been proposed that cover some aspects of either the scientific domain or the provenance information itself, e.g.\,workflow systems~\cite{Belhajjame2008,Altintas2006}, jupyter notebooks~\cite{Samuel2018}, and lab documentation~\cite{Soldatova2014,Giraldo2014}.
In this paper, we discuss a unification of these definitions along typical biomedical research processes, which can be easily adapted to other scientific domains.

Scientific investigations in the biomedical domain can be categorized by their kind into:
\begin{inparaenum}
\item \textit{in-silico},
\item \textit{in-vitro}, and
\item \textit{in-vivo experiments}.
\end{inparaenum}
Often, in-silico studies are performed to simulate real world phenomena based on data obtained from previous in-vitro or in-vivo experiments, i.e. experiments in laboratories respectively living organisms.
New findings from the in-silico studies are then again used to validate the effects in in-vitro or in-vivo experiments, leading to a closed loop of data exchange in this scientific process with respect to the experimental kind.
For the discussion of the particular provenance concepts, we employ the following use case, that we restrict on the two kinds, in-silico and in-vitro, as the challenges and requirements for provenance tracking in in-vitro and in-vivo are quite similar.

\noindent\textbf{\textit{Use case.}} \quad\!
A researcher starts with in-vitro experiments in a wetlab environment measuring the calcium ion mobilization in osteoblasts.
In order to measure this real world phenomenon, a series of actions is performed in the laboratory environment.
The course of action is standardized in a protocol and documented in a so-called \textit{electronic lab notebook (ELN}).
In particular, this involves the preparation of cell environments as well as the set-up and measuring of the prepared cells using a microscope in conjunction with a stimulation device.
Measurement data are collected as microscopy images that are analysed to extract tabular fluorescence intensity data.
In order to exclude environmental and other issues, the experiment is repeated multiple times, resulting in a series of measurements as well as documentation following the same experimental protocol.

\begin{figure}
\centering
\includegraphics[width=.75\textwidth]{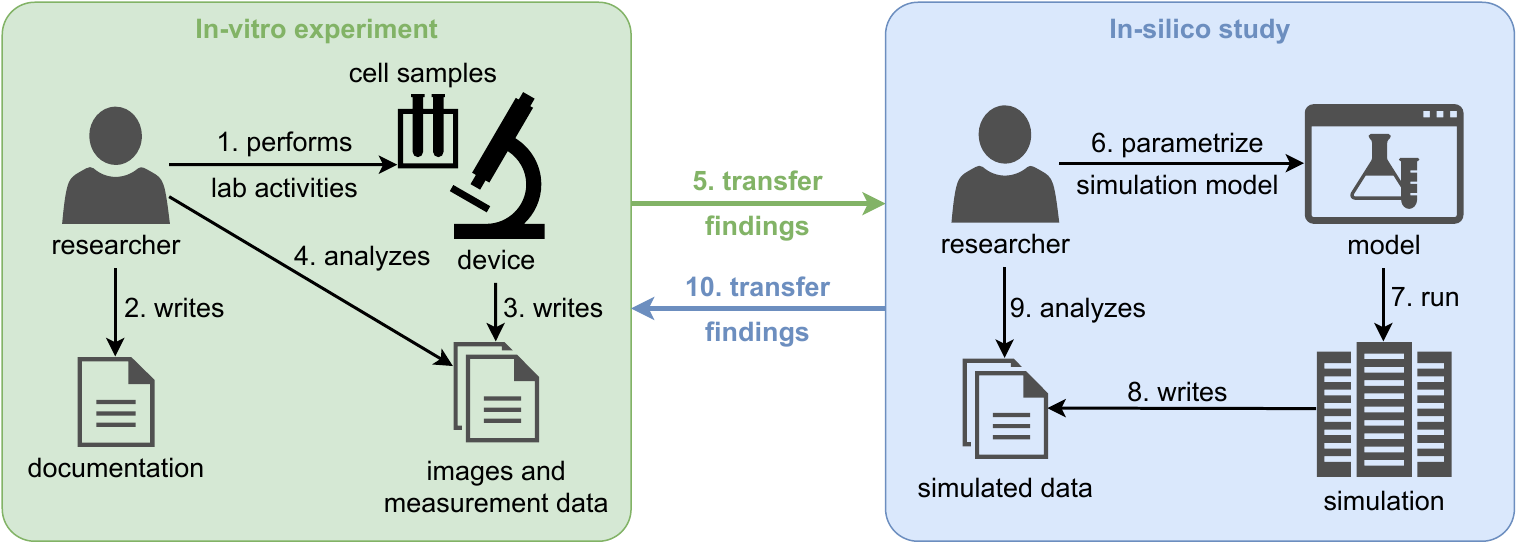}
\caption{Biomedical use case with two experiment types (in-vitro and in-silico) whose research findings are used to validate and optimize the other.
Activities are numbered to illustrate the trajectory.}
\label{fig:use_case}
\end{figure}

In a second step, an in-silico study is set up to further elaborate the stimulation parameter settings based on the real world data from the in-vitro experiments.
The measurements are used to specify and parameterize a simulation model for this phenomenon.
Specifically, the simulation model in combination with the computing environment is run under different simulation settings.
The simulation is repeated in order to reduce random influences.
New findings are then interpreted from all simulations, which provide insights for the optimization of stimulations.
Lastly, these findings are validated in in-vitro experiments again (cf.~Figure~\ref{fig:use_case}).

In this entire sequence of experiments, provenance information is needed to support the credibility and reproducibility of the research findings.
As such, activities, involved persons and entities (e.g.\,data, samples, devices and models) are encoded in the provenance information.

\noindent\textbf{\textit{Contribution.}} \quad\!
\begin{inparaenum}
\item We discuss a unification of workflow and data provenance including the dimension and granularity based on the current state of the art, and
\item we extend the \textbf{W7} provenance question concept~\cite{Ram2007} to \textbf{W7+1} provenance questions and discuss dimensions and granularity for this unification.
\end{inparaenum}

\section{Provenance Concepts}
\label{sec:provenanceConcepts}
Provenance is a very broad term with many meanings and definitions. 
In general, provenance refers to the origin of an object or captures information about the creation or evolution process of an object --- in our example these objects are simulation models and measurement data. 
With respect to the \textit{(scientific) workflow}, provenance information encode the tracing of the particular research objects resulting in the scientific finding.
In the context of a \textit{(scientific) database}~\cite{BT18}, provenance includes information about the origin of a data element and details about its scientific processing (parameters, software versions, etc.). 
Thus, provenance information enable the verification of (scientific) processes as well as database queries~\cite{HDL17,FKSS08,PRS18}.
 
\subsection{Provenance Types}\label{sec:provenanceTypes}
\noindent\textbf{\textit{Workflow provenance.}} \quad\!This type refers to the process of a dataset's derivation in the form of a scientific workflow. 
As described in~\cite{HDL17} a \textit{(scientific) workflow} is a directed graph where nodes represent arbitrary functions or modules in general with some input, output, and parameters. Edges model a predefined data or control flow between these modules. 
It includes information about the workflow's procedure, deviations from it, and the execution in general. 

\begin{figure}[ht]
\centering
\includegraphics[width=\textwidth]{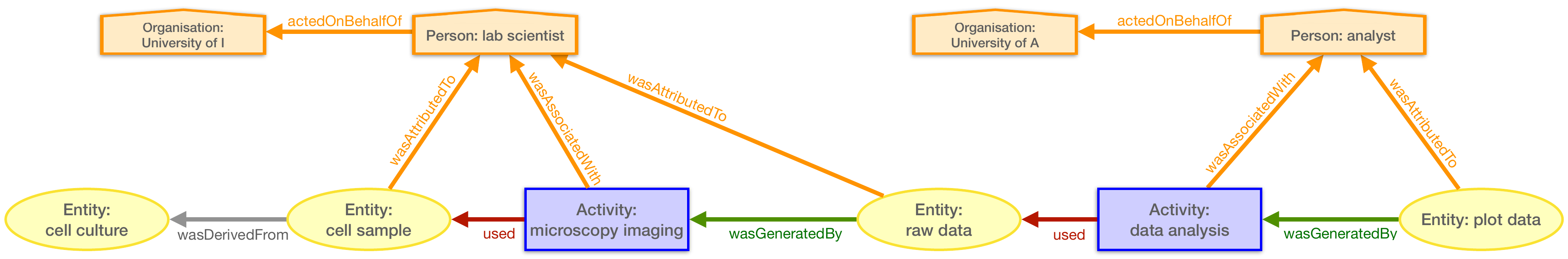}
\caption{Biomedical in-vitro experiment with organizations and persons (\colorbox{PROVorange}{\textcolor{black}{agents}}), cell cultures and samples as well as data sets (\colorbox{PROVyellow}{\textcolor{black}{entities}}), and research activities (\colorbox{PROVblue}{\textcolor{black}{activity}}) including their relationships in the PROV standard. Note that PROV relationship direction is typically from the result to the origin.}
\label{fig:PROV}
\end{figure}

Workflow provenance is often expressed in terms of description logics and derived concepts like ontologies, e.g.\,the PROV-O ontology~\cite{W3C-provPM}. 
The \textit{PROV-O ontology} distinguishes three core concepts: \textit{entities}, \textit{activities}, and \textit{agents}. 
Entities are data or objects and can be derived from other entities. 
Activities can generate or use entities.
Agents can perform or control activities or produce entities. 
An example for our biomedical use case in PROV terms is shown in Figure~\ref{fig:PROV}.

\noindent \textbf{\textit{Data provenance.}}\quad\!
Data provenance plays a role in data wrangling, which can often be expressed in terms of relational algebra.
It describes the derivation of a piece of data from data sets, typically the result of a database query $Q$.
Considering our use case, the following query encodes the comparison of the fluorescence intensity from two stimulation experiments with respect to their voltage: \lstinline{SELECT voltage_2 FROM R NATURAL JOIN S WHERE intensity_1 < intensity_2}, where two tables $R$ and $S$ are joined on a unique join attribute \texttt{sample\_id}, against a source database $D$ (consisting of $R$ and $S$). The query result is stored in a table $T$:
\begin{center}
\scalebox{.575}{
\begin{tabular}{c|c|c|c|cc|c|c|c|cc|c|l}
\cline{2-4} \cline{7-9} \cline{12-12}
$R:$ & \texttt{sample\_id} & \texttt{intensity\_1} & \texttt{voltage}\_1 & & $S:$ & \text{sample\_id} & \texttt{intensity\_2} & \texttt{voltage}\_2 & & $T:$ & \texttt{voltage}\_2 \\
\cline{2-4} \cline{7-9} \cline{12-12}
& 1 & 40.027 & 0.9 & \ \textcolor{RedOrange}{$r_1$} & & 1 & 40.375 & 1.0 & \ \textcolor{RedOrange}{$s_1$} & & 1.0 & \ \textcolor{RedOrange}{$r_1 \cdot s_1 + r_1 \cdot s_3$} \\
\cline{12-12}
& 2 & 41.038 & 1.4 & \ \textcolor{blue}{$r_{2,t_1}$} & & 1 & 39.998 & 1.3 & \ $s_2$ & \multicolumn{1}{c}{} & \multicolumn{1}{c}{} & \ or \textcolor{RedOrange}{$\{\{r_1,s_1\},\{r_1,s_3\}\}$} \\
\multicolumn{1}{c|}{} & 2 & 41.033 & 1.4 & \textcolor{blue}{$r_{2,t_2}$} & & 1 & 42.001 & 1.0 & \ \textcolor{RedOrange}{$s_3$} & \multicolumn{1}{c}{} & \multicolumn{1}{c}{} & \\ 
\cline{2-4} \cline{7-9}
\end{tabular}}
\end{center}
The rows (\textit{tuples}) are provided with \textit{provenance IDs} $r_1$, $s_1$, .... 
In the case of evolving databases (cf.\,Section~\ref{sec:provenanceDimensionsAndGranularity}), additional time stamps are required, processed by \textcolor{blue}{$r_{2,t_1}$}, \textcolor{blue}{$r_{2,t_2}$}, ...

In relational databases, data provenance is often specified in the form of \textit{provenance polynomials}~\cite{GT17} or \textit{witness basis}~\cite{BKT01}.
While the former specifies a concrete calculation rule in the form of a polynomial defined by a commutative semi-ring ${(\mathbb{N}\lbrack X\rbrack,+,\cdot,0,1)}$ with $+$ for duplicates (resulting from projection or union) and $\cdot$ for (natural) joins~\cite{GKT07}, the witness basis describes the set of all relevant witnesses. 
A witness itself contains all the tuple IDs needed to reconstruct a tuple. 
Then, the tuple in the table $T$ can be explained by the provenance polynomial \textcolor{RedOrange}{$r_1 \cdot s_1 + r_1 \cdot s_3$} or the witness basis \textcolor{RedOrange}{$\{\{r_1,s_1\},\{r_1,s_3\}\}$}. 
Both representations contain information about the natural join (\textcolor{RedOrange}{$r_1 \cdot s_1$} or \textcolor{RedOrange}{$\{r_1, s_1\}$}) and the duplicate ($t + t'$ or $\{\{t\}, \{t'\}\}$ for the witnesses $t = \textcolor{RedOrange}{\{r_1,s_1\}}$ and $t' = \textcolor{RedOrange}{\{r_1,s_3\}}$), which are generated by answering the query $Q$.

\subsection{Provenance Dimensions and Granularity}
\textbf{\textit{Dimensions.}}\quad\!
The literature~\cite{HDL17,PRS18} distinguishes between three dimensions of workflow provenance: \textit{retrospective provenance}, \textit{prospective provenance} and \textit{evolution provenance}. 
A provenance solution may support only one, two, or all three of them.
While \textit{retrospective provenance} provides information about past workflow executions and data derivations, \textit{prospective provenance} captures the structure and static context of a workflow~\cite{HDL17}. 
The latter is independent of any workflow execution or input data and can be understood as a recipe for future workflow executions. 
Retrospective provenance, however, preserves information on the resources that are accessed or generated during execution. 
\textit{Evolution provenance} reflects changes made between two iterations of a workflow.

While the three provenance dimensions do not rely on each other's presence~\cite{HDL17}, it may be beneficial to capture multiple dimensions in order to provide detailed information about the executed processes, the underlying procedures, and their relation to other research processes.
For data provenance, to the best of our knowledge, the three dimensions have not yet been discussed.
In Section~\ref{sec:provenanceDimensionsAndGranularity}, we discuss potential applications of dimensions in data provenance.

\noindent\textbf{\textit{Granularity.}} \quad\!
The granularity of a provenance model depends on the level of details with which a workflow is described~\cite{HDL17}.
This level of detail is defined by seven \textit{provenance questions} (cf.\,Section~\ref{sec:provenanceQuestions}).
Depending on how many of the provenance questions~\cite{Ram2007} are answered, the granularity of a model can be considered \textit{coarse-} or \textit{fine-grained}.
While the former yields a broad description of the process, the latter allows for a more detailed description of a process.

\section{Combining the provenance concepts}
\label{sec:combiningConcepts}
Combining data provenance with provenance of workflows provides a more granular insight into research findings by providing insights into the data origin.
This combination is particularly important in biomedical research or data science applications, as the data processing steps are largely carried out outside the (relational) databases; this concerns algorithms for data pre-processing, data cleaning and data transformation.
In order for their effects to be assessable, the workflow provenance information must be added to the data provenance information. 

While workflow provenance typically is encoded in the form of knowledge graphs using W3C PROV~\cite{W3C-provPM}, the data provenance can be encoded within specialized data files (entities) that are also integrated into the workflow model.
Alternatively, the data provenance concept can be extended from tuples in databases to the file level so that data provenance and workflow provenance are specified and processed in parallel but uncoupled (cf.\,Section~\ref{sec:provenanceDimensionsAndGranularity}).

\subsection{Provenance Questions}
\label{sec:provenanceQuestions}
Traceability and reproducibility raise many questions about where things come from and how they were processed:
\textit{Which} datasets are affected by an error or bug? 
\textit{How} are datasets affected by modifying a parameter?
\textit{Why} is an excepted value \textit{not} included (in the result)?
The concept of workflow provenance typically answers seven question types including their combinations (\textbf{W7}, \cite{Ram2007}), whereas data provenance addresses three of these question types: \pq{how}, \pq{why}, and \pq{where}.
However, we propose that also the additional \pq{why not} question, as it is known from \textit{provenance games}~\cite{KLZ13} in the context of data provenance, can be specified on workflow provenance as well as \pq{what}, one of the coarse-grained \textbf{W7} questions, on data provenance.
Though under a closed world assumption, the answering of the remaining negated questions such as \pq{where not} is conceptually valid, but in real world settings infeasible.

Regarding data provenance, answers to \pq{why not} provide an explanation, why an expected result is not part of the query result. 
For example, the query $Q$ from above would, for a value range of $\lbrack 1.0, 1.5\rbrack$ for \texttt{voltage\_2}, never yield the result $1.3$, leading to the insight that the query would to be modified. 
\pq{Why not} regarding workflow provenance has, to the best of our knowledge, not been stated in the literature.
This question, however, contains valuable information to workflow optimization and analysis, as it gives potential reasons about certain choices in experiment design, such as why a particular procedure has been followed instead of another.
We suggest as answers, for instance, notes, design comments, or warnings that are collected during experiment planning and execution.

Schema changes can result in dirty data, including rounding errors or omitting characters (spaces or leading zeros).
To solve this problem, we defined \pq{what} also for data provenance \cite{AH21}. It stores the data type of all attributes.
In workflow provenance, however, \pq{what} could be answered by the order of processes that were performed~\cite{Schroeder2022}.
\pq{When}, \pq{who}, and \pq{which} are only defined for workflow provenance as they are not relevant outside the workflow scope. 

\begin{figure}[ht]
\centering
\includegraphics[width=.85\textwidth]{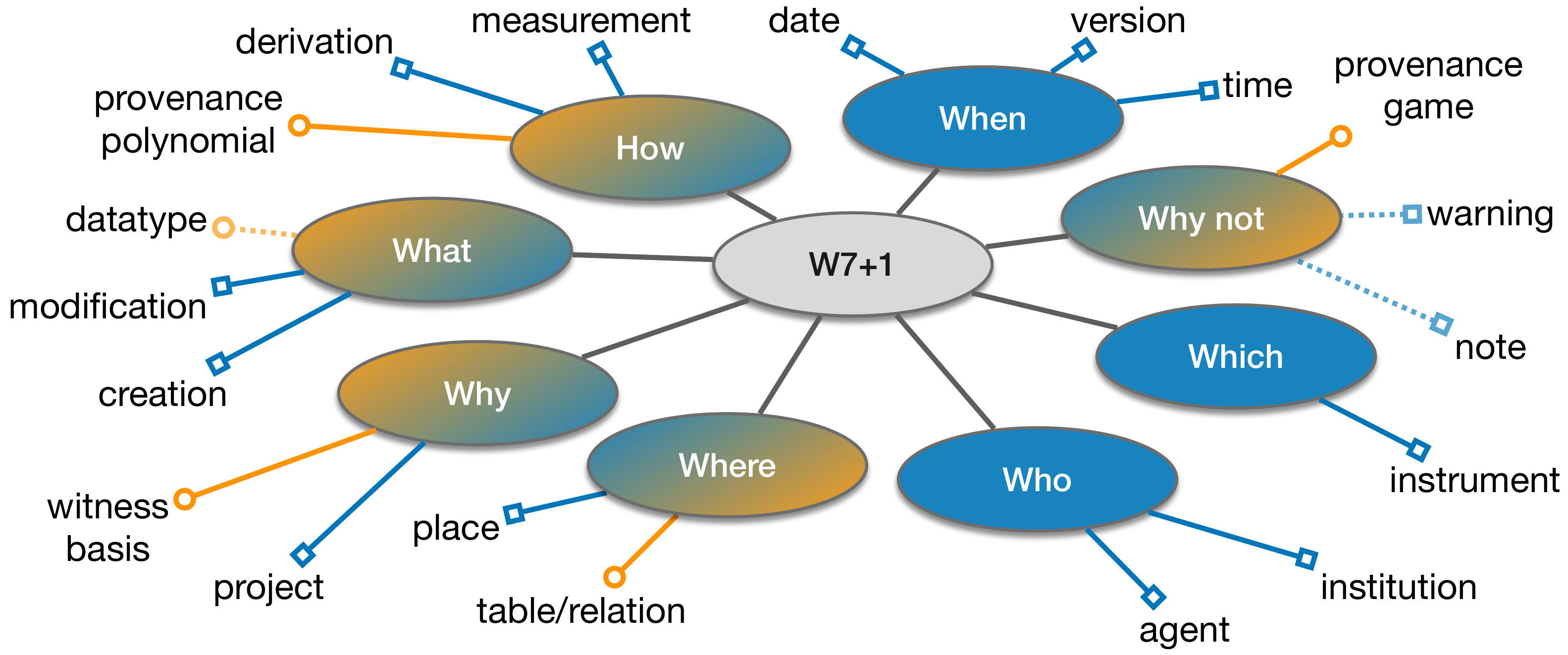}
\caption{\textbf{W7+1} provenance questions for workflow provenance (\textcolor{blueFig}{$\--\!\scalebox{.7}{$\Box$}$}) and data provenance (\textcolor{orange}{$\multimap$}) including typical (solid) and new defined (dotted lines) answer options.}
\label{fig:conceptHierarchy}
\end{figure}

These extensions of the \textbf{W7} questions, we propose to call \textbf{W7+1}.
The \textbf{W7+1} question types as well as their classification according to the provenance types workflow provenance (highlighted in \colorbox{figblue}{\textcolor{white}{blue}}) and data provenance (highlighted in \colorbox{figorange}{\textcolor{black}{orange}}) are shown in Figure~\ref{fig:conceptHierarchy}.

\subsection{Provenance Dimensions and Granularity}
\label{sec:provenanceDimensionsAndGranularity}
Extending the provenance dimensions and granularities towards our unified workflow and data provenance framework, we define them as follows:

\textbf{\textit{Dimensions.}} \quad\!
The three dimensions of provenance with respect to the documentation within wetlabs are discussed in~\cite{Genehr2023}.
Summarizing their arguments, prospective provenance corresponds to the protocol, i.e., the \textit{standard operating procedure (SOP)} in our use case, and retrospective workflow provenance encodes the particular experimental procedure.
Evolution provenance for the workflow comes in, when the simulation results reveal new findings so that the underlying SOP is adjusted for future experiments.

Due to the specification of data provenance reasoning over a specific selection of data, it typically encodes the retrospective dimension.
However, in the case of \textit{evolving databases} such as from measurement data of wetlab experiments including schema changes and data updates over time, the additional dimension of evolution provenance becomes necessary. 
It reflects changes made between two states of a database instances by time stamps $t_1$, $t_2$, ...
As updating data means adding a new modified data record, the clue lies in the choice of the provenance IDs which is suffixed by a time stamp~\cite{AH21}.
The modified tuples receive the same original number but with a different time stamp,
such as \textcolor{blue}{$r_{2,t_1}$} and \textcolor{blue}{$r_{2,t_2}$} in database $D$ of Section~\ref{sec:provenanceTypes}.

\noindent\textbf{\textit{Granularity.}} \quad\!
Depending on the literature, the terms fine- and coarse-grained provenance have different meanings.
While \cite{PRS18} used the former as an alias for data provenance and the latter for workflow provenance, for \cite{HDL17} the granularity definition only exists in the context of workflow provenance, where the coarseness means the level of detail of a workflow description. 

In \cite{Schroeder2022}, the terms fine- and coarse-grained provenance are specified with respect to different levels of granularity for the answers of provenance questions.
Depending on the needs of the user, the question \pq{how} measurement data was created might reflect the entire experiment as a single activity (coarse-grained) or the course of atomic activities in the wetlab (fine-grained).
Finding a balance in the granularity of provenance modelling between fine-grained modelling, which clearly impacts storage and computing resources, and coarse-grained modelling which restricts the expressiveness has previously been identified as one of the core challenges~\cite{Gierend2024}.

In the context of data provenance, we propose the terms to be specified as follows:
While with fine-grained provenance a data element corresponds to the tuples in a database, i.e.\, a particular measurment data point, with coarse-grained provenance a data element corresponds to the entire measurement file, each provided by a unique ID. 
By collecting these provenance IDs in a so-called \textit{ID database} --- in addition to the provenance IDs, the ID database also contains the file name, file paths and further provenance information --- the same evaluations can be performed on the file level as before at the data level.
As such, data provenance and workflow provenance can be combined from a unified viewpoint.

\section{Conclusion}\label{sec:conclusion}
In this paper, we discussed the combination of workflow provenance and data provenance and their effects on the dimensions and granularity employing a biomedical use case.
This discussion represents an initial step towards a unified provenance framework encoding a comprehensive view on research processes by revealing homogeneous concepts within the current literature and necessary extensions.
Thus, this work serves as a motivation for the developing of a formal specification of the unified framework in future.

\begin{acknowledgments}
\textbf{SG} and \textbf{FK} are funded by the Deutsche Forschungsgemeinschaft (DFG, German Research Foundation) - SFB 1270/2 – 299150580. 
\end{acknowledgments}

\section*{Authors’ contributions}
Author contributions according to \href{https://credit.niso.org/}{CRediT}:
\textbf{TA} Conceptualization, Visualization, Writing - Original Draft.
\textbf{SG} Conceptualization, Visualization, Writing - Original Draft.
\textbf{MK} Supervision, Writing - Review \& Editing.
\textbf{FK} Conceptualization, Funding acquisition, Supervision, Writing - Review \& Editing.
\textbf{MS} Conceptualization, Visualization, Writing - Original Draft.
All authors read and approved the final manuscript.

\appendix

\end{document}